\documentclass[11pt]{article}

\usepackage[T1]{fontenc}
\usepackage[utf8]{inputenc}
\usepackage{mathptmx}          
\usepackage[scaled=0.92]{helvet}
\usepackage{courier}
\usepackage{microtype}
\usepackage[letterpaper,margin=1in,headheight=14pt]{geometry}
\usepackage{cite}
\usepackage{amsmath,amssymb}
\usepackage{graphicx}
\usepackage{booktabs}
\usepackage{array}
\usepackage{multirow}
\usepackage{xcolor}
\usepackage{listings}
\usepackage{tikz}
\usetikzlibrary{arrows.meta,positioning,fit,backgrounds}
\usepackage[font=small,labelfont=bf]{caption}
\usepackage[hang,flushmargin]{footmisc}
\usepackage{enumitem}
\usepackage{fancyhdr}
\usepackage{url}
\IfFileExists{xurl.sty}{\usepackage{xurl}}{}
\usepackage[hidelinks]{hyperref}
\hypersetup{
  pdftitle={Containing the Autonomous Operator: A Defense-in-Depth Framework and Reference Architecture for Securing AI Agents on Kubernetes},
  pdfauthor={Simhadri Podala Narasimha},
  pdfkeywords={AI agents, Kubernetes, prompt injection, Model Context Protocol, cloud-native security}
}

\setlist{itemsep=2pt,topsep=4pt}
\fancypagestyle{firstpage}{\fancyhf{}\fancyfoot[C]{\thepage}}

\definecolor{codebg}{gray}{0.96}
\definecolor{codecomment}{gray}{0.45}
\lstdefinelanguage{yaml}{
  keywords={true,false,null},
  sensitive=false,
  comment=[l]{\#},
  morestring=[b]",
  morestring=[b]'
}
\lstdefinelanguage{rego}{
  morekeywords={package,import,default,if,in,contains,some,every,not,with,as},
  sensitive=true,
  comment=[l]{\#},
  morestring=[b]"
}
\newcommand{\prim}{$\bullet$}
\newcommand{\supp}{$\circ$}

\begin{document}

\thispagestyle{firstpage}
\begin{center}
  \rule{\linewidth}{1.5pt}\\[0.9em]
  {\LARGE\bfseries Containing the Autonomous Operator:\\[0.25em]
   A Defense-in-Depth Framework and Reference Architecture\\[0.25em]
   for Securing AI Agents on Kubernetes\par}
  \vspace{0.7em}
  \rule{\linewidth}{0.6pt}\\[1.4em]
  {\large Simhadri Podala Narasimha}\\[0.3em]
  Independent Researcher\\
  Littleton, Colorado, USA\\
  \href{mailto:podalas@gmail.com}{\texttt{podalas@gmail.com}}\\[1em]
  {\small October 2026}
\end{center}
\vspace{0.5em}

\begin{abstract}
\noindent Large language model (LLM) agents are moving from chat interfaces into infrastructure operations, where they read telemetry, call tools, generate and execute code, and change the state of production Kubernetes clusters. This collapses a boundary that conventional cloud-native security assumes: the boundary between data and control. Content that an agent merely reads---a log line, a ticket, a tool description---can redirect what it does. This paper argues that the model must not be treated as a security boundary and that agent safety on Kubernetes is therefore an infrastructure problem: every guarantee must continue to hold under the assumption that the agent is fully compromised by prompt injection. We contribute (i)~a threat model and ten-class threat taxonomy for agents operating on and within Kubernetes, aligned with emerging OWASP guidance for agentic applications; (ii)~nine design principles, centered on complete mediation at the tool boundary and on breaking the combination of untrusted input, sensitive access, and external egress; (iii)~a seven-layer defense-in-depth framework that maps each principle to native or widely adopted Kubernetes mechanisms---workload identity, RBAC and ValidatingAdmissionPolicy, gVisor/Kata sandboxing via the SIG Apps Agent Sandbox project, FQDN-aware egress policy, an agent/MCP gateway with policy-as-code over tool arguments, and eBPF runtime enforcement; (iv)~a reference architecture with concrete policy artifacts and per-layer bindings for Amazon EKS, Azure Kubernetes Service, and Google Kubernetes Engine; and (v)~a qualitative evaluation comprising a threat--control coverage matrix and four attack walkthroughs, with a proposed empirical methodology. We report no measured attack-success or overhead figures; instead we identify residual risks and the measurements needed to validate the framework.

\vspace{0.6em}
\end{abstract}

\vspace{0.5em}
\noindent{\footnotesize\itshape The views expressed are the author's own and do not represent those of any employer.}

\section{Introduction}
\label{sec:intro}

Operations teams are beginning to delegate real work to LLM-based agents: triaging alerts, reading logs and metrics, proposing and applying remediations, opening pull requests against GitOps repositories, and generating ad hoc scripts to diagnose problems. Kubernetes is a natural home for these agents. It is where the workloads they manage already run, its declarative API is well suited to automated reasoning, and an ecosystem of agent frameworks, protocols, and gateways has grown around it, including the Model Context Protocol (MCP)~\cite{mcp-spec}, the Agent2Agent (A2A) protocol, the CNCF kagent project~\cite{kagent}, and the Linux Foundation agentgateway project~\cite{agentgateway}.

An agent that can act on a cluster is, from a security standpoint, a new kind of principal. Unlike a controller, its behavior is not fixed by code that can be reviewed: it is a function of a model, a prompt, and every piece of content that enters its context window during a task. Unlike a human operator, it does not reliably distinguish instructions from data. Indirect prompt injection---adversarial instructions embedded in content the agent retrieves rather than in the user's request---has been demonstrated against deployed LLM-integrated applications~\cite{greshake2023}, and benchmark studies show that tool-using agents remain susceptible even with model-level defenses~\cite{agentdojo}. In the MCP ecosystem specifically, \emph{tool poisoning}---hidden instructions in tool descriptions, schemas, or return values---has been demonstrated against widely used clients~\cite{invariant-tpa, owasp-mcp-cheatsheet}.

For a Kubernetes operator this has a sharp consequence. Content that an agent merely \emph{reads}---a container log line written by an attacker-controlled workload, an annotation on a resource, a field in an incident ticket, a tool description served by a third-party MCP server---can become a command that the agent then executes with its own credentials. The classical separation between the data plane and the control plane, on which much of cloud-native security rests, does not survive contact with an agent unless the platform reconstructs it.

This paper takes the position that \emph{the model is not a security boundary}. Model alignment, system prompts, and input classifiers reduce the probability of misbehavior, but they offer no guarantee, and their failure modes are adversarially discoverable. Agent security on Kubernetes must therefore be built the way multi-tenant platform security is built: by assuming the workload is hostile and constraining what it can reach, do, and exfiltrate, regardless of what it ``intends.'' This framing aligns with the classical principles of least privilege and complete mediation~\cite{saltzer1975} and with the confused-deputy problem~\cite{hardy1988}, both of which return with new force when the deputy is an LLM.

\subsection{Contributions}
\begin{enumerate}
  \item A \textbf{threat model} for LLM agents that operate \emph{on} a Kubernetes cluster (as operators) and \emph{within} it (as workloads), with a ten-class taxonomy aligned with OWASP guidance for agentic applications~\cite{owasp-agentic-tm, owasp-agentic-top10} (Section~\ref{sec:threat}).
  \item Nine \textbf{design principles} that make agent safety independent of model behavior (Section~\ref{sec:principles}).
  \item A seven-layer \textbf{defense-in-depth framework} that realizes the principles with Kubernetes-native and widely adopted open-source mechanisms, with concrete policy artifacts (Section~\ref{sec:framework}).
  \item A \textbf{reference architecture} with graduated autonomy tiers and a per-layer mapping onto the three major managed Kubernetes services---Amazon EKS, Azure Kubernetes Service (AKS), and Google Kubernetes Engine (GKE)---including where each provides native controls and where the gaps must be filled with open-source components (Section~\ref{sec:arch}).
  \item A \textbf{qualitative evaluation}: a threat--control coverage matrix, four attack walkthroughs, an analysis of residual risk, and a proposed methodology for empirical validation (Section~\ref{sec:eval}).
\end{enumerate}

\subsection{Scope and Non-Goals}
We focus on agents used for infrastructure and platform operations---the setting in which agent actions have the largest blast radius. We do not address model training-time attacks (data or weight poisoning), the safety of model outputs presented to humans, or hardening of the inference service itself. We treat the model provider as reachable over the network and outside the trust boundary of the cluster. The policy artifacts in this paper are illustrative; API versions of fast-moving projects should be checked against current releases.

\section{Background}
\label{sec:background}

\subsection{LLM Agents}
An LLM agent iterates a loop: it receives a goal, reasons over its context, selects a tool and arguments, observes the result, and repeats until it decides the goal is met~\cite{react}. Three properties matter for security. First, \emph{tool results re-enter the context}, so data returned by any tool can influence subsequent tool choices. Second, \emph{tool selection is open-ended}: the agent chooses which capability to invoke and with what arguments at run time. Third, agents increasingly carry \emph{memory} across sessions---vector stores, scratchpads, or summaries---so a single injection can persist.

\subsection{Agent Protocols}
MCP standardizes how an agent (client) discovers and invokes tools, resources, and prompts exposed by servers~\cite{mcp-spec}. Tool descriptions and schemas are natural-language metadata that the model reads to decide when and how to call a tool. Because this metadata is consumed by the model rather than by deterministic code, it is an injection surface in its own right~\cite{invariant-tpa, hou2025mcp, sok-mcp}. A2A standardizes agent-to-agent messaging and capability discovery. Both protocols are now commonly fronted by gateways that terminate transport security, authenticate callers, and apply per-tool authorization~\cite{agentgateway}.

\subsection{Kubernetes Security Primitives}
Kubernetes provides a rich but fragmented set of controls: ServiceAccounts and projected, audience-bound tokens for workload identity; RBAC for coarse authorization by verb and resource; admission control, including the in-process, CEL-based ValidatingAdmissionPolicy (VAP)~\cite{k8s-vap}, for fine-grained validation of object content; Pod Security Standards; \texttt{RuntimeClass} for selecting alternative container runtimes such as gVisor~\cite{gvisor} or Kata Containers~\cite{kata}; NetworkPolicy and CNI-specific extensions such as Cilium's FQDN-aware policies~\cite{cilium}; and API-server audit logging~\cite{k8s-audit}. eBPF-based tools such as Tetragon~\cite{tetragon} and Falco~\cite{falco} add kernel-level observation and enforcement. Managed Kubernetes services add workload-identity federation that maps ServiceAccounts to short-lived cloud credentials: EKS Pod Identity and IAM Roles for Service Accounts (IRSA) on AWS~\cite{eks-pod-identity}, Microsoft Entra Workload ID on AKS~\cite{aks-wid}, and Workload Identity Federation for GKE~\cite{gke-wif}. They also increasingly offer managed sandbox runtimes and FQDN-aware egress policy, which we compare in Section~\ref{sec:arch}.

None of these mechanisms was designed with agents in mind, and none, alone, addresses the data/control collapse described above. The contribution of this paper is to compose them, together with agent-specific components, into a coherent posture.

\section{Threat Model}
\label{sec:threat}

\subsection{System Model}
We consider an \emph{operations agent} deployed as a workload in a Kubernetes cluster (or in a management cluster that administers workload clusters). The agent receives tasks from humans (chat, tickets) or from automation (alerts). It reasons using a hosted or self-hosted model, calls tools exposed over MCP (Kubernetes API reads, log and metrics queries, GitOps operations, cloud-provider APIs), and may request \emph{code-execution sandboxes} in which it runs generated scripts. It may store memory in a vector database, and it may delegate sub-tasks to other agents over A2A.

\subsection{Assets}
The assets at risk are: (a)~the integrity and availability of cluster workloads and configuration; (b)~credentials reachable from the agent or its tools (ServiceAccount tokens, cloud credentials, API keys); (c)~confidential data visible through tools (logs, configuration, customer data in telemetry); (d)~the integrity of the agent's memory; and (e)~the audit trail that attributes actions to principals.

\subsection{Adversary Capabilities}
We consider five adversary capabilities, which may combine:
\begin{itemize}
  \item[\textbf{A1}] \textbf{Content injection.} The adversary controls some content the agent will read: a log line emitted by a workload they control, an annotation or label, a ticket body, a web page, a document in a retrieval corpus.
  \item[\textbf{A2}] \textbf{Malicious or compromised tool server.} The adversary operates an MCP server the agent connects to, or compromises an existing one, and can change its tool metadata or responses at any time.
  \item[\textbf{A3}] \textbf{Malicious generated code.} As a consequence of A1 or A2, code the agent writes and executes is attacker-influenced.
  \item[\textbf{A4}] \textbf{Low-privilege insider.} A user permitted to make requests of the agent attempts to use it to exceed their own authority.
  \item[\textbf{A5}] \textbf{Benign failure.} No adversary is present, but model error, hallucination, or a feedback loop produces harmful actions.
\end{itemize}
We assume the adversary \emph{can} fully control the agent's subsequent tool choices and arguments once injection succeeds. This is the central modeling choice: we grant the attacker the agent.

\subsection{Trust Assumptions}
We trust the Kubernetes control plane, the node kernel and container runtime (subject to the isolation strength of the selected \texttt{RuntimeClass}), the policy engines, the gateway, and the identity provider. We do not trust the model, the agent process, its context, retrieved content, tool metadata from servers outside an approved registry, or generated code.

\subsection{Threat Taxonomy}
Table~\ref{tab:threats} lists ten threat classes. The taxonomy is aligned in spirit with the OWASP Agentic AI threats-and-mitigations work and the OWASP Top~10 for Agentic Applications~\cite{owasp-agentic-tm, owasp-agentic-top10}, specialized here to Kubernetes.

\begin{table}[t]
\centering
\caption{Threat taxonomy for agents on Kubernetes}
\label{tab:threats}
\renewcommand{\arraystretch}{1.15}
\footnotesize
\begin{tabular}{@{}p{0.08\linewidth}p{0.27\linewidth}p{0.55\linewidth}@{}}
\toprule
\textbf{ID} & \textbf{Threat} & \textbf{Kubernetes manifestation} \\
\midrule
T1 & Goal hijack via indirect prompt injection & A crafted log line or annotation instructs the agent to delete, scale, or reconfigure a workload. \\
T2 & Tool misuse / excessive agency & The agent invokes a legitimate tool with harmful arguments (e.g., \texttt{replicas: 0} in production). \\
T3 & Privilege compromise / confused deputy & The agent acts with its own broad ServiceAccount instead of the requesting user's authority. \\
T4 & Tool and MCP supply chain & Tool poisoning, rug-pull redefinition, tool shadowing, or a typosquatted MCP server image. \\
T5 & Untrusted code execution and escape & Generated code exploits a kernel vulnerability or reads a mounted token from a shared-kernel container. \\
T6 & Data and credential exfiltration & The agent is induced to send secrets or logs to an attacker-controlled endpoint. \\
T7 & Memory and context poisoning & Injected content is written to long-term memory and influences future sessions. \\
T8 & Cascading or runaway action & A remediation loop restarts or rescales repeatedly; fan-out exhausts quotas or budget. \\
T9 & Inter-agent spoofing & A rogue agent impersonates a trusted peer over A2A, or relays injected instructions. \\
T10 & Repudiation and untraceability & Actions cannot be attributed to a requester, prompt, and approval. \\
\bottomrule
\end{tabular}
\end{table}

\subsection{The Lethal Combination}
A useful lens is the observation, popularized by Willison~\cite{willison-trifecta} and formalized as the ``Agents Rule of Two'' by Meta~\cite{meta-rule-of-two}, that serious harm typically requires an agent session to combine three properties at once: exposure to untrusted input, access to sensitive data or privileged actions, and the ability to communicate externally or change state. Removing any one of the three for a given session sharply bounds the damage of a successful injection. Kubernetes is unusually well suited to enforcing this structurally, because each property maps onto a controllable resource: content sources onto tool authorization, sensitive access onto RBAC and secret placement, and external communication onto egress policy.

\section{Design Principles}
\label{sec:principles}

\textbf{P1 -- The model is not a security boundary.} Every control must hold if the agent is fully adversary-controlled. Prompt-level defenses are hygiene, not guarantees.

\textbf{P2 -- Complete mediation at the tool boundary.} Every action the agent takes leaves the agent process through a mediated channel---the gateway or the Kubernetes API server---where it is authenticated, authorized against its \emph{arguments}, and logged. There is no unmediated path.

\textbf{P3 -- Least agency.} Grant the minimum set of tools, verbs, resources, and namespaces needed for the task class, and prefer read-only capabilities. Agency should be scoped per task, not per agent product.

\textbf{P4 -- Delegated, attributable identity.} Each agent has its own workload identity; actions on behalf of a user carry that user's delegated authority, so the agent cannot exceed the requester's permissions.

\textbf{P5 -- Isolate untrusted execution.} Generated code runs in ephemeral, kernel-isolated sandboxes with no ambient credentials.

\textbf{P6 -- Break the lethal combination.} Deny egress by default, and arrange sessions so that untrusted input, sensitive access, and external communication do not coexist without human approval.

\textbf{P7 -- Prefer reversible, reviewable actions.} Route state changes through GitOps pull requests where possible, so they gain review, history, and rollback.

\textbf{P8 -- Observable and attributable by construction.} Correlate each request, model call, tool call, API-server action, and approval in a single trace.

\textbf{P9 -- Bounded autonomy.} Rate-limit actions, cap budgets, and trip circuit breakers; autonomy is graduated by environment and by action risk.

\section{A Seven-Layer Defense-in-Depth Framework}
\label{sec:framework}

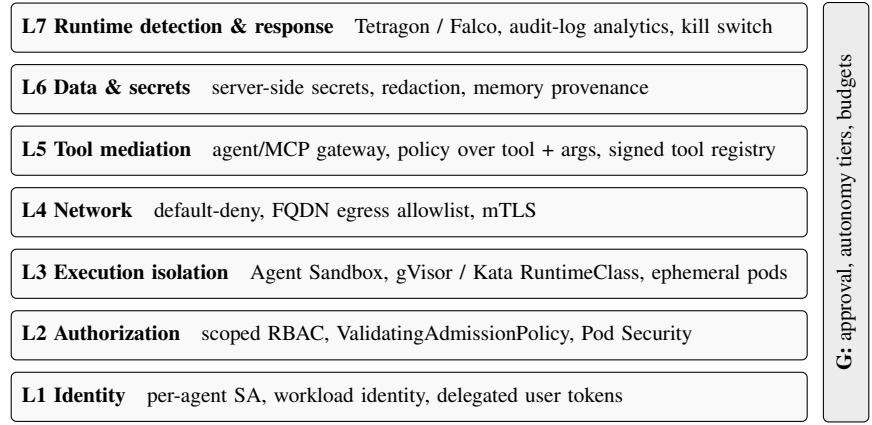
\begin{figure}[t]
\centering
\begin{tikzpicture}[
  node distance=1.5mm,
  layer/.style={draw, rounded corners=2pt, text width=0.62\linewidth, minimum height=6.5mm, align=left, font=\scriptsize, inner xsep=4pt},
  gov/.style={draw, rounded corners=2pt, minimum height=6.5mm, font=\footnotesize, align=center, fill=gray!15}
]
\node[layer, fill=gray!5]  (l7) {\textbf{L7 Runtime detection \& response}\quad Tetragon / Falco, audit-log analytics, kill switch};
\node[layer, fill=gray!5, below=of l7] (l6) {\textbf{L6 Data \& secrets}\quad server-side secrets, redaction, memory provenance};
\node[layer, fill=gray!5, below=of l6] (l5) {\textbf{L5 Tool mediation}\quad agent/MCP gateway, policy over tool + args, signed tool registry};
\node[layer, fill=gray!5, below=of l5] (l4) {\textbf{L4 Network}\quad default-deny, FQDN egress allowlist, mTLS};
\node[layer, fill=gray!5, below=of l4] (l3) {\textbf{L3 Execution isolation}\quad Agent Sandbox, gVisor / Kata RuntimeClass, ephemeral pods};
\node[layer, fill=gray!5, below=of l3] (l2) {\textbf{L2 Authorization}\quad scoped RBAC, ValidatingAdmissionPolicy, Pod Security};
\node[layer, fill=gray!5, below=of l2] (l1) {\textbf{L1 Identity}\quad per-agent SA, workload identity, delegated user tokens};
\node[gov, fit=(l1)(l7), inner sep=0pt, right=2mm of l4.east, anchor=west, minimum width=6mm, text width=4mm] (g) {};
\node[font=\scriptsize, rotate=90] at (g.center) {\textbf{G:} approval, autonomy tiers, budgets};
\end{tikzpicture}
\caption{Seven technical layers (L1--L7) and a cross-cutting governance layer (G).}
\label{fig:layers}
\end{figure}

Fig.~\ref{fig:layers} summarizes the framework. Layers L1--L7 are technical controls; a cross-cutting governance layer (G) covers approval, autonomy tiers, and budgets.

\subsection{L1: Identity}
\label{sec:l1}

\textbf{Per-agent workload identity.} Every agent, and every class of tool server, runs under its own ServiceAccount. The default ServiceAccount is never used, and \texttt{automountServiceAccountToken} is disabled on any pod that does not call the Kubernetes API, notably on code-execution sandboxes. Where tokens are needed, projected tokens should be audience-bound and short-lived. Cloud access uses workload identity---EKS Pod Identity or IRSA~\cite{eks-pod-identity}, Microsoft Entra Workload ID~\cite{aks-wid}, or Workload Identity Federation for GKE~\cite{gke-wif}---rather than static keys, so credentials are scoped, rotated, and attributable in the provider's audit trail (AWS CloudTrail, Azure Monitor and Entra sign-in logs, or Cloud Audit Logs). For mesh-wide identity across clusters and non-Kubernetes services, SPIFFE/SPIRE~\cite{spiffe} provides attestable workload identities usable for mutual TLS.

\textbf{Delegated user authority.} To address the confused-deputy threat (T3), an agent acting on behalf of a user should present the user's delegated authority, not only its own. OAuth~2.0 Token Exchange~\cite{rfc8693} allows the gateway to exchange the user's token for a down-scoped token that carries both the user (\texttt{sub}) and the agent (\texttt{act}) as actors. Tool servers then authorize against the intersection of the two. On the Kubernetes API itself, the agent can use user impersonation headers, permitted by a tightly scoped \texttt{impersonate} RBAC rule that restricts which users or groups may be impersonated, so that RBAC is evaluated against the human's permissions and the audit log records both identities.

\subsection{L2: Authorization and Admission}
\label{sec:l2}

\textbf{RBAC.} RBAC answers \emph{which verbs on which resources in which namespaces}. Agents receive namespaced Roles, never ClusterRoles with write verbs. Read-only roles (\texttt{get}, \texttt{list}, \texttt{watch}) are the default; \texttt{pods/exec}, \texttt{secrets}, \texttt{serviceaccounts/token}, and RBAC objects are excluded, since each allows a direct path to privilege escalation.

\textbf{Admission as an agent guardrail.} RBAC cannot express constraints on object \emph{content}: it cannot forbid scaling to zero while permitting scaling, or forbid an agent from editing objects in control-plane namespaces while permitting edits elsewhere. ValidatingAdmissionPolicy~\cite{k8s-vap} fills this gap with in-process CEL expressions that can be scoped to requests from agent identities via \texttt{matchConditions}. Listing~\ref{lst:vap} shows a policy that applies only to ServiceAccounts in the \texttt{agents} namespace. Because it is enforced by the API server, it holds even if the agent bypasses the gateway and calls the API directly with a stolen token---a property that tool-layer controls alone cannot provide. Kyverno or OPA Gatekeeper can express equivalent rules where richer policy libraries are needed.

\begin{lstlisting}[float=tbp,language=yaml, caption={ValidatingAdmissionPolicy restricting agent-originated changes.}, label={lst:vap}]
apiVersion: admissionregistration.k8s.io/v1
kind: ValidatingAdmissionPolicy
metadata:
  name: agent-guardrails
spec:
  failurePolicy: Fail
  matchConstraints:
    resourceRules:
    - apiGroups: ["apps"]
      apiVersions: ["v1"]
      operations: ["CREATE", "UPDATE", "DELETE"]
      resources: ["deployments", "deployments/scale", "statefulsets", "statefulsets/scale"]
  matchConditions:
  - name: only-agent-identities
    expression: "request.userInfo.username.startsWith('system:serviceaccount:agents:')"
  validations:
  - expression: "!request.namespace.startsWith('kube-') && request.namespace != 'agents'"
    message: "agents may not modify control-plane or agent namespaces"
  - expression: "request.operation != 'DELETE'"
    message: "agents may not delete workloads; open a GitOps PR instead"
  - expression: "request.operation == 'DELETE' || !has(object.spec.replicas) || object.spec.replicas > 0"
    message: "agents may not scale workloads to zero"
---
apiVersion: admissionregistration.k8s.io/v1
kind: ValidatingAdmissionPolicyBinding
metadata:
  name: agent-guardrails-binding
spec:
  policyName: agent-guardrails
  validationActions: ["Deny", "Audit"]
\end{lstlisting}

\textbf{Pod Security.} Namespaces that host agents and sandboxes enforce the \texttt{restricted} Pod Security Standard, which prevents privileged pods, host namespaces, and hostPath mounts---the most direct paths from a compromised pod to the node.

\subsection{L3: Execution Isolation}
\label{sec:l3}

Agents that execute generated code (T5) need stronger isolation than a standard container, because containers share the host kernel and a single kernel vulnerability is sufficient for escape. Two mature options are available through \texttt{RuntimeClass}: gVisor~\cite{gvisor}, which interposes a user-space kernel that services system calls on behalf of the container and thereby shrinks the host-kernel surface, at a cost in syscall-heavy workloads~\cite{young2019gvisor}; and Kata Containers~\cite{kata}, which runs each pod in a lightweight virtual machine with its own guest kernel.

The Kubernetes SIG Apps \textbf{Agent Sandbox} project~\cite{agent-sandbox} provides a \texttt{Sandbox} custom resource and controller for exactly this use case: isolated, stateful, singleton workloads with a stable identity, intended among other things for running untrusted, LLM-generated code. Its extensions add templates, claims, and pre-warmed pools to reduce start-up latency, which matters because per-session sandboxes are only practical if they are fast to obtain. Listing~\ref{lst:sandbox} shows a hardened sandbox: kernel-isolated runtime, no ServiceAccount token, non-root, read-only root filesystem, all capabilities dropped, a digest-pinned image, and resource limits.

\begin{lstlisting}[float=tbp,language=yaml, caption={Hardened code-execution sandbox (API group/version per project release).}, label={lst:sandbox}]
apiVersion: agents.x-k8s.io/v1alpha1
kind: Sandbox
metadata:
  name: session-7f3a
  namespace: agent-sandboxes
  labels: {app.kubernetes.io/component: agent-sandbox}
spec:
  podTemplate:
    spec:
      runtimeClassName: gvisor   # GKE Sandbox; kata-vm-isolation on AKS
      automountServiceAccountToken: false
      securityContext:
        runAsNonRoot: true
        seccompProfile: {type: RuntimeDefault}
      containers:
      - name: runtime
        image: registry.internal/sandbox/python@sha256:<digest>
        securityContext:
          allowPrivilegeEscalation: false
          readOnlyRootFilesystem: true
          capabilities: {drop: ["ALL"]}
        resources:
          limits: {cpu: "1", memory: 1Gi, ephemeral-storage: 512Mi}
        volumeMounts:
        - {name: scratch, mountPath: /workspace}
      volumes:
      - name: scratch
        emptyDir: {sizeLimit: 512Mi}
\end{lstlisting}

Sandboxes should be ephemeral and per session; a sandbox that persists across users becomes a channel for cross-session contamination (T7). They are placed in a dedicated namespace with no RBAC bindings, and their network access is governed by L4.

\subsection{L4: Network}
\label{sec:l4}

Network policy is the most direct way to break the exfiltration leg of the lethal combination (T6). Agent and sandbox namespaces start from default-deny for both ingress and egress. Egress is then allowed only to cluster DNS, to the agent gateway, and to an explicit allowlist of external hostnames---typically the model endpoint. Because external services are addressed by name and their IP addresses change, IP-based NetworkPolicy is insufficient; FQDN-aware policies, as provided by Cilium~\cite{cilium}, are required. Listing~\ref{lst:cilium} permits an agent to reach only DNS, the gateway, and a single model endpoint. Sandboxes typically receive no external egress at all; when a task genuinely needs internet access, egress should be through an inspecting proxy and should trigger the approval path (Section~\ref{sec:gov}).

\begin{lstlisting}[float=tbp,language=yaml, caption={FQDN-restricted egress for agent pods (Cilium).}, label={lst:cilium}]
apiVersion: cilium.io/v2
kind: CiliumNetworkPolicy
metadata:
  name: agent-egress
  namespace: agents
spec:
  endpointSelector:
    matchLabels: {app.kubernetes.io/component: agent}
  egress:
  - toEndpoints:
    - matchLabels:
        k8s:io.kubernetes.pod.namespace: kube-system
        k8s-app: kube-dns
    toPorts:
    - ports: [{port: "53", protocol: ANY}]
      rules:
        dns: [{matchPattern: "*"}]
  - toEndpoints:
    - matchLabels:
        k8s:io.kubernetes.pod.namespace: agentgateway-system
    toPorts:
    - ports: [{port: "8080", protocol: TCP}]
  - toFQDNs:
    - matchName: bedrock-runtime.us-west-2.amazonaws.com
    toPorts:
    - ports: [{port: "443", protocol: TCP}]
\end{lstlisting}

The same intent is expressed with provider-native resources on AKS and GKE (Section~\ref{sec:providers}, Listing~\ref{lst:gkefqdn}). Note that the agent pod has no route to the Kubernetes API server in this policy. All cluster actions go through tool servers behind the gateway, which hold the necessary credentials. This realizes P2: the agent's only path to state change is mediated. East--west traffic between gateway, tool servers, and agents should use mutual TLS, either from a service mesh or from SPIFFE-issued certificates.

\subsection{L5: Tool Mediation}
\label{sec:l5}

The tool boundary is where agent intent becomes action, and it is the most important point of control. We place an \emph{agent gateway} between agents and all tool servers (MCP), peer agents (A2A), and, optionally, model endpoints. agentgateway~\cite{agentgateway}, now a Linux Foundation project, is one open-source realization; it supports MCP and A2A, includes MCP- and A2A-aware authorization, and integrates with Kubernetes through the Gateway API via kgateway. The gateway performs five functions.

\textbf{(1) Authentication.} It authenticates the calling agent by workload identity and the originating user by token, and performs the token exchange described in Section~\ref{sec:l1}.

\textbf{(2) Authorization over tool and arguments.} Per-tool allow lists are necessary but insufficient (T2): the harm in \texttt{scale(deployment, 0)} lies in the argument, not the tool name. The gateway therefore sends a structured decision request---agent, user, tool, arguments, session attributes---to a policy engine such as OPA~\cite{opa} or Cedar. Listing~\ref{lst:rego} shows a Rego policy that permits read tools within an agent's namespaces, permits GitOps pull requests only to approved repositories, requires human approval for disruptive actions in production, and denies scale-to-zero outright. Admission policy (L2) enforces overlapping constraints at the API server; the duplication is intentional, since L5 can block earlier with richer context while L2 cannot be bypassed.

\begin{lstlisting}[float=tbp,language=rego, caption={Policy over tool calls and their arguments (OPA/Rego v1).}, label={lst:rego}]
package agent.tools

import rego.v1

read_only := {"k8s_get", "k8s_list", "k8s_logs", "prom_query"}
disruptive := {"k8s_scale", "k8s_rollout_restart", "k8s_cordon"}

default decision := {"allow": false, "approval": false}

agent := data.agents[input.agent]

in_scope if input.args.namespace in agent.namespaces

deny contains "scale-to-zero is forbidden" if {
  input.tool == "k8s_scale"
  input.args.replicas == 0
}

deny contains "cross-namespace access" if {
  input.args.namespace
  not in_scope
}

decision := {"allow": true, "approval": false} if {
  count(deny) == 0
  input.tool in read_only
  in_scope
}

decision := {"allow": true, "approval": false} if {
  count(deny) == 0
  input.tool == "gitops_open_pr"
  input.args.repo in agent.repos
}

decision := {"allow": true, "approval": true} if {
  count(deny) == 0
  input.tool in disruptive
  in_scope
  startswith(input.args.namespace, "prod-")
}
\end{lstlisting}

\textbf{(3) Supply-chain control for tools (T4).} Tool servers are admitted only from an internal registry, with images signed and verified at admission (e.g., Sigstore/cosign~\cite{sigstore} with a Kyverno or policy-controller rule). Tool descriptions and schemas are pinned: the gateway records a hash of each tool's metadata at approval time and rejects or quarantines a server whose metadata changes---the ``rug-pull'' pattern identified in OWASP's MCP guidance~\cite{owasp-mcp-cheatsheet}. Tools with colliding or confusable names across servers (shadowing) are rejected at registration.

\textbf{(4) Response handling.} Tool responses are untrusted input. The gateway can strip markup commonly used to smuggle instructions, enforce size limits, and tag provenance so the agent framework can render tool output as quoted data. These measures reduce but do not eliminate injection; they are hygiene under P1, not a boundary.

\textbf{(5) Logging and rate limiting.} Every call is logged with its decision and correlated trace ID (L7, G), and rate limits per agent, tool, and session bound runaway behavior (T8).

\textbf{Risk-tiered tools.} We classify tools into three tiers: \emph{observe} (reads; auto-execute), \emph{propose} (open a PR, create a dry-run plan; auto-execute, since a human reviews the result), and \emph{act} (direct mutation; approval-gated in production). Exposing a ``propose'' alternative for every ``act'' tool lets most remediation flow through GitOps (P7), where Argo~CD or Flux applies changes with history and rollback.

\subsection{L6: Data and Secrets}
\label{sec:l6}

\textbf{Secrets never enter the context.} Credentials are held by tool servers, fetched server-side from a secret manager (e.g., via the External Secrets Operator or Vault), and never returned in tool output. An agent that never sees a secret cannot be induced to leak it, which removes an entire class of exfiltration regardless of egress policy.

\textbf{Redaction at the source.} Log and event tools redact tokens, keys, and personal data before returning results, both to protect confidentiality and to avoid feeding sensitive material to an external model.

\textbf{Memory provenance (T7).} Writes to long-term memory record their source (user, tool, session) and trust level. Content derived from untrusted tool output is not promoted to memory without review, memory is partitioned per tenant and per agent, and memory stores support point-in-time rollback so a poisoning incident can be reversed.

\textbf{Information-flow separation.} The strongest available defense against T1 is architectural: separate the component that reads untrusted data from the component that plans privileged actions, so that untrusted content can influence data values but not control flow. The CaMeL design~\cite{camel} demonstrates this pattern with a privileged planner and a quarantined reader, enforced by capability tracking in an interpreter. On Kubernetes, the pattern maps naturally onto separate pods with separate identities and tool grants: a quarantined ``reader'' agent with read-only tools and no egress, and a ``planner'' agent that never ingests raw untrusted content and whose actions pass through L5.

\subsection{L7: Runtime Detection and Response}
\label{sec:l7}

Preventive layers will be misconfigured or bypassed; detection bounds the dwell time.

\textbf{Kernel-level enforcement.} eBPF tools observe agent and sandbox pods at the system-call level. Tetragon~\cite{tetragon} can both detect and enforce, for example terminating a process that attempts to open a ServiceAccount token path inside a sandbox that should have none (Listing~\ref{lst:tetragon}). Falco~\cite{falco} offers a broad rule set for unexpected shells, network connections, and file writes.

\begin{lstlisting}[float=tbp,language=yaml, caption={Tetragon policy terminating token access in sandboxes.}, label={lst:tetragon}]
apiVersion: cilium.io/v1alpha1
kind: TracingPolicy
metadata:
  name: sandbox-token-guard
spec:
  podSelector:
    matchLabels: {app.kubernetes.io/component: agent-sandbox}
  kprobes:
  - call: "security_file_open"
    syscall: false
    args:
    - {index: 0, type: "file"}
    selectors:
    - matchArgs:
      - index: 0
        operator: "Prefix"
        values: ["/var/run/secrets/kubernetes.io", "/var/run/secrets/eks.amazonaws.com"]
      matchActions:
      - action: Sigkill
\end{lstlisting}

\textbf{Control-plane analytics.} API-server audit logs~\cite{k8s-audit}, filtered to agent identities, are baselined: new verbs, new namespaces, or abnormal call rates raise alerts. Gateway decision logs supply denied-call spikes, which are a strong signal of an injection attempt in progress.

\textbf{Kill switch.} A single, rehearsed action must disable an agent: revoke or rotate its ServiceAccount and workload-identity bindings, scale it to zero, apply a quarantine network policy, and invalidate its gateway sessions. Because the agent's only paths are mediated (P2), revoking identity at L1 and L5 is sufficient to stop it even if its pod persists.

\subsection{G: Governance---Approval, Autonomy, and Budgets}
\label{sec:gov}

\textbf{Human approval.} Approval-gated calls are suspended at the gateway and surfaced to a human with the full context: requester, rationale, exact arguments, and a dry-run diff. Approval is bound to the specific call (by hash of tool and arguments) and expires, so it cannot be replayed for a different action.

\textbf{Autonomy tiers.} Autonomy is graduated by environment (Table~\ref{tab:tiers}). An agent may be fully autonomous for reads in production and for most actions in development, while production mutations require approval or GitOps review.

\textbf{Budgets and circuit breakers.} Each agent and session has budgets for tool calls, mutations per time window, model tokens, and spend. A circuit breaker opens after repeated failed remediations of the same object, after repeated policy denials, or when a mutation does not improve the triggering signal, and hands the incident to a human. This directly targets runaway loops (T8), which are the most likely harm in practice because they require no adversary (A5).

\begin{table}[t]
\centering
\caption{Graduated autonomy tiers}
\label{tab:tiers}
\renewcommand{\arraystretch}{1.15}
\footnotesize
\setlength{\tabcolsep}{3pt}
\begin{tabular}{@{}lccc@{}}
\toprule
\textbf{Tool tier} & \textbf{Dev} & \textbf{Staging} & \textbf{Production} \\
\midrule
Observe (read)        & auto & auto & auto \\
Propose (PR, dry-run) & auto & auto & auto$^{\dagger}$ \\
Act (mutate)          & auto & notify & approval \\
Destructive (delete)  & approval & approval & denied (L2) \\
\bottomrule
\end{tabular}\\[2pt]
{\scriptsize $^{\dagger}$A human reviews and merges the pull request.}
\end{table}

\section{Reference Architecture on Managed Kubernetes: EKS, AKS, and GKE}
\label{sec:arch}

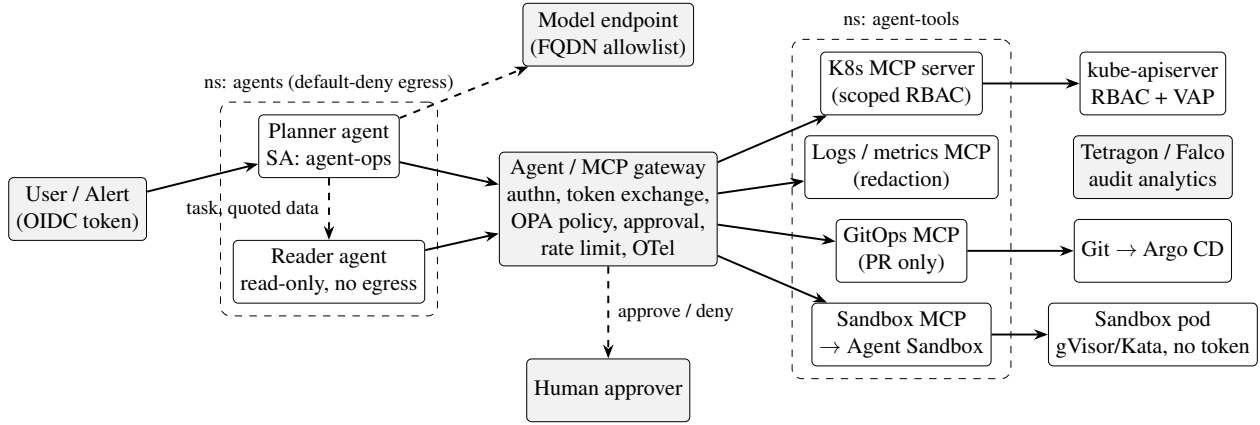
\begin{figure}[t]
\centering
\resizebox{\linewidth}{!}{%
\begin{tikzpicture}[
  font=\footnotesize,
  box/.style={draw, rounded corners=2pt, align=center, minimum height=9mm, inner sep=3pt, fill=white},
  ns/.style={draw, dashed, rounded corners=4pt, inner sep=5pt},
  arr/.style={-{Stealth[length=2mm]}, thick},
  darr/.style={-{Stealth[length=2mm]}, thick, dashed}
]
\node[box, fill=gray!10] (user) at (0,0) {User / Alert\\(OIDC token)};
\node[box] (planner) at (3.6,0.9) {Planner agent\\SA: agent-ops};
\node[box] (reader) at (3.6,-0.9) {Reader agent\\read-only, no egress};
\node[box, fill=gray!10, minimum width=26mm] (gw) at (7.6,0) {Agent / MCP gateway\\authn, token exchange,\\OPA policy, approval,\\rate limit, OTel};
\node[box] (k8stool) at (11.8,1.8) {K8s MCP server\\(scoped RBAC)};
\node[box] (obs) at (11.8,0.6) {Logs / metrics MCP\\(redaction)};
\node[box] (git) at (11.8,-0.6) {GitOps MCP\\(PR only)};
\node[box] (sbx) at (11.8,-1.8) {Sandbox MCP\\$\rightarrow$ Agent Sandbox};
\node[box] (api) at (15.4,1.8) {kube-apiserver\\RBAC + VAP};
\node[box] (argo) at (15.4,-0.6) {Git $\rightarrow$ Argo CD};
\node[box] (sand) at (15.4,-1.8) {Sandbox pod\\gVisor/Kata, no token};
\node[box, fill=gray!10] (llm) at (7.6,2.5) {Model endpoint\\(FQDN allowlist)};
\node[box, fill=gray!10] (hitl) at (7.6,-2.6) {Human approver};
\node[box, fill=gray!10] (det) at (15.4,0.6) {Tetragon / Falco\\audit analytics};

\begin{scope}[on background layer]
\node[ns, fit=(planner)(reader), label={[font=\scriptsize]above:ns: agents (default-deny egress)}] {};
\node[ns, fit=(k8stool)(obs)(git)(sbx), label={[font=\scriptsize]above:ns: agent-tools}] {};
\end{scope}

\draw[arr] (user) -- (planner);
\draw[arr] (planner) -- (gw);
\draw[arr] (reader) -- (gw);
\draw[darr] (planner) -- node[left,font=\scriptsize]{task, quoted data} (reader);
\draw[darr] (planner) -- (llm);
\draw[arr] (gw) -- (k8stool);
\draw[arr] (gw) -- (obs);
\draw[arr] (gw) -- (git);
\draw[arr] (gw) -- (sbx);
\draw[arr] (k8stool) -- (api);
\draw[arr] (git) -- (argo);
\draw[arr] (sbx) -- (sand);
\draw[darr] (gw) -- node[right,font=\scriptsize]{approve / deny} (hitl);
\end{tikzpicture}}
\caption{Provider-neutral reference architecture (provider bindings in Table~\ref{tab:providers}). Agent pods have no route to the API server or the internet except the model endpoint; every action is mediated by the gateway, and mutations are re-validated by admission control.}
\label{fig:arch}
\end{figure}

Fig.~\ref{fig:arch} shows a reference deployment. Its core---namespaces, RBAC, admission policy, the agent gateway and its policy engine, GitOps, and eBPF runtime enforcement---is provider-neutral and runs unchanged on Amazon EKS, Azure Kubernetes Service (AKS), and Google Kubernetes Engine (GKE). Five concerns bind to the provider: cloud workload identity, the kernel-isolated sandbox runtime, FQDN-aware egress enforcement, the managed model endpoint, and managed detection and audit services. Section~\ref{sec:providers} maps each layer onto the three services.

\subsection{Components}
\begin{itemize}
  \item \textbf{Agent namespace.} A planner agent and a quarantined reader agent (Section~\ref{sec:l6}), each with its own ServiceAccount federated, through the provider's workload identity, to a narrowly scoped cloud role that can invoke only the model endpoint. Default-deny egress with an FQDN allowlist. No Kubernetes API access.
  \item \textbf{Gateway.} An MCP/A2A gateway that authenticates agents and users, performs token exchange, calls OPA for per-call decisions, holds approval-gated calls, enforces rate limits and budgets, and emits OpenTelemetry spans.
  \item \textbf{Tool namespace.} Signed, digest-pinned MCP servers, each with its own ServiceAccount and minimal RBAC: a Kubernetes server with read and narrowly scoped write roles, an observability server that redacts, a GitOps server that can only open pull requests, and a sandbox server that requests per-session \texttt{Sandbox} objects.
  \item \textbf{Sandbox namespace.} Agent Sandbox controller with gVisor or Kata \texttt{RuntimeClass}, warm pools, no RBAC bindings, no egress.
  \item \textbf{Control plane guardrails.} VAP bindings matched to agent and tool identities; Pod Security \texttt{restricted} on all three namespaces; image-signature verification at admission.
  \item \textbf{Detection.} Tetragon policies on agent and sandbox pods; audit-log pipeline with per-identity baselines; gateway decision logs; a scripted kill switch.
\end{itemize}

\subsection{Provider Bindings}
\label{sec:providers}

Table~\ref{tab:providers} maps each framework layer onto the native services of EKS, AKS, and GKE. Three differences matter most for agent workloads.

\begin{table}[t]
\centering
\caption{Mapping of framework layers onto managed Kubernetes services. Entries reflect provider documentation as of late 2026; availability varies by region and cluster version.}
\label{tab:providers}
\renewcommand{\arraystretch}{1.2}
\footnotesize
\begin{tabular}{@{}>{\raggedright\arraybackslash}p{0.12\linewidth}>{\raggedright\arraybackslash}p{0.26\linewidth}>{\raggedright\arraybackslash}p{0.26\linewidth}>{\raggedright\arraybackslash}p{0.26\linewidth}@{}}
\toprule
\textbf{Layer} & \textbf{Amazon EKS} & \textbf{Azure AKS} & \textbf{Google GKE} \\
\midrule
L1 Workload identity & EKS Pod Identity or IRSA $\rightarrow$ IAM role & Microsoft Entra Workload ID (federated credentials) & Workload Identity Federation for GKE $\rightarrow$ IAM \\
L2 Admission & VAP; Kyverno or Gatekeeper & VAP; Azure Policy for AKS (Gatekeeper-based) & VAP; Policy Controller \\
L2 Image trust & Image signing with AWS Signer/Notation, verified at admission & Notation signatures on ACR, verified with Ratify & Binary Authorization \\
L3 Sandbox runtime & No managed option; self-installed gVisor or Kata \texttt{RuntimeClass} on suitable nodes & Pod Sandboxing: Kata VM isolation (\texttt{kata-vm-isolation}) on Azure Linux node pools & GKE Sandbox (gVisor) \\
L3 Sandbox orchestration & Open-source Agent Sandbox & Open-source Agent Sandbox (Kata on AKS example) & Managed GKE Agent Sandbox add-on (warm pools, claims) \\
L4 FQDN egress & Not supported by VPC CNI network policy; Cilium or an egress proxy required & ACNS FQDN filtering with Azure CNI powered by Cilium & \texttt{FQDNNetworkPolicy} with GKE Dataplane V2 \\
L4 Private model access & PrivateLink (VPC endpoints) to Bedrock & Private Endpoint to Azure OpenAI & Private Service Connect; VPC Service Controls \\
L5 Gateway and policy & \multicolumn{3}{c}{Provider-neutral: agentgateway (or equivalent) with OPA/Cedar; identical on all three} \\
L6 Secrets & Secrets Manager via Secrets Store CSI or External Secrets & Key Vault provider for Secrets Store CSI & Secret Manager add-on or External Secrets \\
L6 Model endpoint & Amazon Bedrock & Azure OpenAI & Vertex AI \\
L7 Managed detection & GuardDuty EKS Runtime Monitoring & Microsoft Defender for Containers & Security Command Center Container Threat Detection \\
L7 Audit & EKS control-plane logs; CloudTrail & AKS diagnostic logs to Azure Monitor & Cloud Audit Logs \\
\bottomrule
\end{tabular}
\end{table}

\textbf{Sandboxing is the largest gap.} GKE offers the most complete managed path: GKE Agent Sandbox is a managed add-on built on the open-source Agent Sandbox controller, uses gVisor for kernel-level isolation, and adds pre-warmed pools for sub-second provisioning, a claim-based request model, and default-deny networking for sandboxes~\cite{gke-agent-sandbox}. AKS provides managed Kata-based Pod Sandboxing, in which each pod runs in a lightweight VM with its own guest kernel, selected with \texttt{runtimeClassName: kata-vm-isolation}; it requires Azure Linux node pools~\cite{aks-pod-sandboxing}, and the open-source Agent Sandbox project documents its use on AKS~\cite{agent-sandbox-aks}. On EKS there is no managed sandbox runtime at the time of writing, so operators install gVisor or Kata themselves on dedicated node groups and run the open-source controller. That is feasible but places the isolation runtime, and its patching, inside the operator's responsibility.

\textbf{FQDN egress is native on AKS and GKE, not on EKS.} On AKS, FQDN filtering is part of Advanced Container Networking Services (ACNS) and requires Azure CNI powered by Cilium; it is a paid add-on, and Microsoft notes possible degradation above roughly 1{,}000 requests per second per pod~\cite{aks-fqdn}. On GKE, the \texttt{FQDNNetworkPolicy} resource requires GKE Dataplane V2 and caps each FQDN at 50 resolved addresses~\cite{gke-fqdn}; Listing~\ref{lst:gkefqdn} expresses the intent of Listing~\ref{lst:cilium}. On EKS, the VPC CNI's network-policy support is not FQDN-aware, so the egress allowlist of Section~\ref{sec:l4} requires Cilium (in chaining or replacement mode) or a forced egress proxy. On all three, private connectivity to the model endpoint (PrivateLink, Private Endpoint, or Private Service Connect) lets the allowlist name a private address rather than a public one, and on GKE, VPC Service Controls adds a perimeter around Vertex AI and other Google APIs that limits where data can flow even with valid credentials.

\begin{lstlisting}[float=tbp,language=yaml, caption={GKE equivalent of Listing~\ref{lst:cilium}: egress limited to the model endpoint (GKE Dataplane V2). A standard NetworkPolicy additionally permits DNS and the gateway.}, label={lst:gkefqdn}]
apiVersion: networking.gke.io/v1alpha1
kind: FQDNNetworkPolicy
metadata:
  name: agent-model-egress
  namespace: agents
spec:
  podSelector:
    matchLabels: {app.kubernetes.io/component: agent}
  egress:
  - matches:
    - name: "us-central1-aiplatform.googleapis.com"
    ports:
    - {protocol: "TCP", port: 443}
\end{lstlisting}

\textbf{Identity semantics differ in detail.} All three federate a Kubernetes ServiceAccount to a cloud principal without static keys, which satisfies L1. They differ in where the binding is declared---an EKS Pod Identity association or IRSA role trust policy, an Entra federated identity credential, or a Google IAM policy binding on the Kubernetes principal---and therefore in which team controls it. In every case the binding should be owned by the platform team and managed declaratively (for example with Crossplane or Terraform), so that an agent with write access to its own namespace cannot widen its cloud permissions by editing ServiceAccount annotations.

\textbf{Portability.} Because the gateway, policy engine, admission policies, and Tetragon policies are provider-neutral, an organization running agents across clouds can keep a single policy repository and vary only the provider overlay---the \texttt{RuntimeClass} name, the egress-policy resource, the identity binding, and the model endpoint. We regard this split as a design goal: the controls that encode \emph{what an agent may do} should not depend on the cloud, while the controls that encode \emph{how isolation is achieved} may.

\subsection{Request Walkthrough}
An alert reports elevated error rates for \texttt{checkout} in \texttt{prod-payments}. (1)~The alert carries a service identity; the planner receives the task. (2)~The planner asks the reader to summarize recent logs and metrics; the reader calls the observability tools through the gateway, which authorizes read-only access to \texttt{prod-payments}, and returns a structured summary to the planner as quoted data. (3)~The planner proposes a rollout restart. The gateway's policy classifies \texttt{k8s\_rollout\_restart} in a \texttt{prod-} namespace as approval-gated and suspends the call. (4)~An on-call engineer approves the exact call; the gateway forwards it with an exchanged token. (5)~The Kubernetes tool server applies the change; the API server evaluates RBAC and the VAP against the request. (6)~Every step is a span in one trace, from alert to API-server audit event.

\section{Evaluation}
\label{sec:eval}

We evaluate the framework qualitatively, through threat coverage and adversarial walkthroughs, and we propose a methodology for empirical validation. We do not report attack-success rates, latency, or cost overheads; those require the experiments described in Section~\ref{sec:method}.

\subsection{Threat--Control Coverage}
Table~\ref{tab:coverage} maps each threat to the layers that provide primary (\prim) or supporting (\supp) mitigation. Two properties are worth noting. First, every threat has at least two mitigating layers, and for every threat at least one primary control is enforced outside the agent process; no threat relies solely on model behavior (P1). Second, the injection-driven threats (T1, T2, T6) are each covered by a preventive control at the tool boundary (L5) \emph{and} an independent control at a different enforcement point (L2 at the API server, L4 in the CNI), so a single misconfiguration does not lead to harm.

\begin{table}[t]
\centering
\caption{Threat--control coverage (\prim\ primary, \supp\ supporting)}
\label{tab:coverage}
\renewcommand{\arraystretch}{1.15}
\setlength{\tabcolsep}{3pt}
\footnotesize
\begin{tabular}{@{}l*{8}{c}@{}}
\toprule
\textbf{Threat} & \textbf{L1} & \textbf{L2} & \textbf{L3} & \textbf{L4} & \textbf{L5} & \textbf{L6} & \textbf{L7} & \textbf{G} \\
\midrule
T1 Goal hijack            & \supp & \prim &       & \supp & \prim & \prim & \supp & \prim \\
T2 Tool misuse            &       & \prim &       &       & \prim &       & \supp & \prim \\
T3 Confused deputy        & \prim & \prim &       &       & \supp &       & \supp &       \\
T4 Tool supply chain      &       & \supp & \supp & \supp & \prim &       & \supp & \supp \\
T5 Code exec / escape     & \supp &       & \prim & \supp &       &       & \prim &       \\
T6 Exfiltration           & \supp &       &       & \prim & \supp & \prim & \supp &       \\
T7 Memory poisoning       & \supp &       & \supp &       & \supp & \prim &       & \supp \\
T8 Runaway action         &       & \supp &       &       & \prim &       & \prim & \prim \\
T9 Inter-agent spoofing   & \prim &       &       & \supp & \prim &       &       &       \\
T10 Repudiation           & \prim &       &       &       & \prim &       & \supp & \prim \\
\bottomrule
\end{tabular}
\end{table}

\subsection{Attack Walkthroughs}

\textbf{S1: Injection through a log line (T1, T2).} An attacker who controls a workload in \texttt{prod-payments} writes a log line instructing any reading agent to ``free capacity by scaling \texttt{checkout} to zero.'' In a naive deployment with a broadly privileged agent, this succeeds if the model follows the instruction. In the reference architecture, the log is read by the quarantined reader, whose output reaches the planner as quoted data (L6); if the planner nonetheless proposes \texttt{k8s\_scale} with \texttt{replicas: 0}, the gateway policy denies it (L5, Listing~\ref{lst:rego}); were the gateway misconfigured, the VAP rejects the request at the API server (L2, Listing~\ref{lst:vap}); and the cluster of denials raises an alert (L7). The attack must defeat three independent enforcement points to cause harm.

\textbf{S2: Tool poisoning with exfiltration (T4, T6).} A third-party MCP server, approved months earlier, updates its tool description to instruct the agent to read a configuration secret and include it in a parameter of the server's own tool. The gateway detects that the server's tool-metadata hash has changed and quarantines it pending review (L5). Independently, the agent cannot read secrets---no tool returns them and no RBAC grants them (L2, L6)---and the server's pod, under default-deny egress, cannot reach an arbitrary external host (L4). The combination of untrusted input and external communication exists in the compromised server, but the sensitive-access leg is absent.

\textbf{S3: Malicious generated code (T5).} Injected content causes the agent to generate a script that attempts to read a ServiceAccount token and then exploit a kernel vulnerability. The sandbox has no token mounted (L1); Tetragon terminates the process on access to the token path (L7, Listing~\ref{lst:tetragon}); the gVisor or Kata runtime interposes a separate kernel boundary, so a host-kernel exploit must also defeat the user-space or guest kernel (L3); and with no egress, the sandbox cannot fetch a second-stage payload or exfiltrate results (L4). The residual risk is a vulnerability in the isolation runtime itself.

\textbf{S4: Runaway remediation (T8).} With no adversary, the agent repeatedly restarts a deployment whose failures are caused by a downstream database, worsening the incident. Per-object mutation budgets and the circuit breaker (G) stop the loop after a bounded number of attempts and escalate to a human; gateway rate limits (L5) cap the rate in the meantime; and audit-log baselines flag the burst (L7).

\subsection{Residual Risk}
Several risks are reduced but not eliminated.
\begin{itemize}
  \item \textbf{Harmful but policy-compliant actions.} Policy can forbid known-bad arguments; it cannot judge whether a permitted action is wise. A restart that is allowed may still be wrong. Approval gates and GitOps review move this judgment to humans, at the cost of latency and of approval fatigue.
  \item \textbf{Integrity attacks within scope.} An injected agent can still produce misleading summaries or propose subtly harmful pull requests. Review quality, not infrastructure, is the control.
  \item \textbf{Exfiltration through the model channel.} The model endpoint is, by necessity, an allowed egress destination. Sensitive data in the context is sent to the model provider by design; contractual and data-residency controls apply, and L6 minimizes what enters the context.
  \item \textbf{Isolation-runtime and policy-engine vulnerabilities.} These components join the trusted computing base.
  \item \textbf{Policy drift.} The layered design produces many policies across tools; their consistency must itself be tested.
\end{itemize}

\subsection{Proposed Methodology for Empirical Validation}
\label{sec:method}
To move from qualitative to quantitative claims, we propose the following experiments, which we leave to future work.

\begin{enumerate}
  \item \textbf{Kubernetes-native injection benchmark.} Adapt the structure of AgentDojo~\cite{agentdojo}---paired user tasks and injection tasks in a realistic environment---to a cluster environment, with injections placed in logs, events, annotations, ConfigMaps, tickets, and MCP tool metadata. Measure benign task utility and targeted attack success rate (ASR) under ablations: no controls; prompt-level defenses only; each layer added in isolation; and the full stack.
  \item \textbf{Bypass testing.} For each enforcement point, disable the layers above it and confirm the lower layer alone prevents the harmful outcome, verifying the independence claimed in Table~\ref{tab:coverage}.
  \item \textbf{Overhead.} Measure added latency per tool call at the gateway and policy engine, sandbox acquisition time with and without warm pools, and syscall-heavy workload slowdown under gVisor and Kata, building on prior characterizations~\cite{young2019gvisor}.
  \item \textbf{Operator burden.} Measure approval volume and approval latency per incident under each autonomy tier, as a proxy for approval fatigue.
\end{enumerate}

\section{Discussion}
\label{sec:discussion}

\textbf{Why not rely on better models?} Model-level robustness to injection is improving, and design-level defenses such as CaMeL~\cite{camel} show that strong guarantees are possible when control and data flow are separated. Infrastructure controls are complementary: they bound the damage of the residual failures, they apply uniformly across models and frameworks, and they are already familiar to platform teams. A platform that assumes compromise remains safe when the model is swapped.

\textbf{Utility trade-offs.} Every layer constrains the agent. Default-deny egress breaks tools that fetch documentation; approval gates slow remediation; quarantined readers lose nuance. We view the autonomy tiers as the primary dial: organizations can begin with observe-and-propose in production and expand autonomy per action class as evidence accumulates.

\textbf{Agents as tenants.} The framework is, in effect, multi-tenant platform engineering with the agent as the untrusted tenant. Much existing practice---namespace isolation, admission control, network segmentation, runtime detection---transfers directly. What is new is the tool boundary, where authorization must consider arguments and provenance, and the need to treat natural-language metadata as code.

\textbf{Standardization gaps.} Several needed primitives lack standards today: a common schema for tool-call authorization requests, signed and versioned tool metadata in MCP, a standard for propagating delegated user identity through multi-agent chains, and conventions for tracing agent reasoning across OpenTelemetry spans. Progress on these would make the controls here portable across gateways and frameworks.

\textbf{Limitations.} This work is a framework and design study. Its evaluation is analytical rather than empirical; the coverage matrix reflects our assessment rather than measurement; and the artifacts, while concrete, have not been subjected to a red-team exercise reported here. The ecosystem is moving quickly, and specific projects or API versions cited may change.

\section{Related Work}
\label{sec:related}

\textbf{Prompt injection.} Greshake et al.\ introduced indirect prompt injection against LLM-integrated applications~\cite{greshake2023}. AgentDojo provides a dynamic benchmark of attacks and defenses for tool-using agents~\cite{agentdojo}. CaMeL defends by design, separating control and data flow with capability tracking~\cite{camel}. Our L6 information-flow separation adopts this idea at the level of pods and identities.

\textbf{Agent and MCP security.} OWASP has published threats and mitigations for agentic AI~\cite{owasp-agentic-tm}, a Top~10 for agentic applications~\cite{owasp-agentic-top10}, and an MCP security cheat sheet~\cite{owasp-mcp-cheatsheet}, alongside its LLM Top~10~\cite{owasp-llm-top10}. The Cloud Security Alliance's MAESTRO framework offers a layered threat-modeling method for agentic systems~\cite{maestro}. Tool poisoning was disclosed by Invariant Labs~\cite{invariant-tpa}, and surveys systematize MCP threats~\cite{hou2025mcp, sok-mcp}. These works characterize threats and general mitigations; we contribute a concrete mapping onto Kubernetes enforcement points and a reference architecture.

\textbf{Risk management.} The NIST AI Risk Management Framework~\cite{nist-ai-rmf} provides organizational structure for AI risk; our governance layer is one operational instantiation for infrastructure agents.

\textbf{Cloud-native isolation and policy.} Sandboxed runtimes~\cite{gvisor, kata, young2019gvisor}, admission control~\cite{k8s-vap}, policy as code~\cite{opa}, eBPF runtime security~\cite{tetragon, falco}, workload identity~\cite{spiffe}, and software signing~\cite{sigstore} are established. The Agent Sandbox project~\cite{agent-sandbox}, kagent~\cite{kagent}, and agentgateway~\cite{agentgateway} bring agent-specific constructs to Kubernetes. Our contribution is the composition of these into a principled posture whose guarantees hold under full agent compromise.

\section{Conclusion}
\label{sec:conclusion}

AI agents turn content into commands. On Kubernetes, where the cost of a wrong command is measured in outages, the safe response is to treat the agent as an untrusted tenant and rebuild the data/control boundary in infrastructure. We presented a threat model, nine principles, a seven-layer framework with concrete Kubernetes artifacts, and a reference architecture in which every action is mediated, every guarantee survives full agent compromise, and autonomy is earned per action class. The most important next step is empirical: a Kubernetes-native injection benchmark that measures utility and attack success under layered ablations, so that platform teams can choose their autonomy tiers on evidence.


\end{document}